\documentclass[letterpaper,compsoc,twoside]{IEEEtran}
\usepackage{fixltx2e} 
\usepackage{cmap} 
\usepackage{ifthen}
\usepackage[T1]{fontenc}
\usepackage[utf8]{inputenc}
\usepackage{amsmath}

\usepackage[font={small,it},labelfont=bf]{caption}
\usepackage{float}

\pdfoutput=1
\usepackage{scipy}
\makeatletter
\def\PY@reset{\let\PY@it=\relax \let\PY@bf=\relax%
    \let\PY@ul=\relax \let\PY@tc=\relax%
    \let\PY@bc=\relax \let\PY@ff=\relax}
\def\PY@tok#1{\csname PY@tok@#1\endcsname}
\def\PY@toks#1+{\ifx\relax#1\empty\else%
    \PY@tok{#1}\expandafter\PY@toks\fi}
\def\PY@do#1{\PY@bc{\PY@tc{\PY@ul{%
    \PY@it{\PY@bf{\PY@ff{#1}}}}}}}
\def\PY#1#2{\PY@reset\PY@toks#1+\relax+\PY@do{#2}}

\expandafter\def\csname PY@tok@gd\endcsname{\def\PY@tc##1{\textcolor[rgb]{0.63,0.00,0.00}{##1}}}
\expandafter\def\csname PY@tok@gu\endcsname{\let\PY@bf=\textbf\def\PY@tc##1{\textcolor[rgb]{0.50,0.00,0.50}{##1}}}
\expandafter\def\csname PY@tok@gt\endcsname{\def\PY@tc##1{\textcolor[rgb]{0.00,0.27,0.87}{##1}}}
\expandafter\def\csname PY@tok@gs\endcsname{\let\PY@bf=\textbf}
\expandafter\def\csname PY@tok@gr\endcsname{\def\PY@tc##1{\textcolor[rgb]{1.00,0.00,0.00}{##1}}}
\expandafter\def\csname PY@tok@cm\endcsname{\let\PY@it=\textit\def\PY@tc##1{\textcolor[rgb]{0.25,0.50,0.56}{##1}}}
\expandafter\def\csname PY@tok@vg\endcsname{\def\PY@tc##1{\textcolor[rgb]{0.73,0.38,0.84}{##1}}}
\expandafter\def\csname PY@tok@m\endcsname{\def\PY@tc##1{\textcolor[rgb]{0.13,0.50,0.31}{##1}}}
\expandafter\def\csname PY@tok@mh\endcsname{\def\PY@tc##1{\textcolor[rgb]{0.13,0.50,0.31}{##1}}}
\expandafter\def\csname PY@tok@cs\endcsname{\def\PY@tc##1{\textcolor[rgb]{0.25,0.50,0.56}{##1}}\def\PY@bc##1{\setlength{\fboxsep}{0pt}\colorbox[rgb]{1.00,0.94,0.94}{\strut ##1}}}
\expandafter\def\csname PY@tok@ge\endcsname{\let\PY@it=\textit}
\expandafter\def\csname PY@tok@vc\endcsname{\def\PY@tc##1{\textcolor[rgb]{0.73,0.38,0.84}{##1}}}
\expandafter\def\csname PY@tok@il\endcsname{\def\PY@tc##1{\textcolor[rgb]{0.13,0.50,0.31}{##1}}}
\expandafter\def\csname PY@tok@go\endcsname{\def\PY@tc##1{\textcolor[rgb]{0.20,0.20,0.20}{##1}}}
\expandafter\def\csname PY@tok@cp\endcsname{\def\PY@tc##1{\textcolor[rgb]{0.00,0.44,0.13}{##1}}}
\expandafter\def\csname PY@tok@gi\endcsname{\def\PY@tc##1{\textcolor[rgb]{0.00,0.63,0.00}{##1}}}
\expandafter\def\csname PY@tok@gh\endcsname{\let\PY@bf=\textbf\def\PY@tc##1{\textcolor[rgb]{0.00,0.00,0.50}{##1}}}
\expandafter\def\csname PY@tok@ni\endcsname{\let\PY@bf=\textbf\def\PY@tc##1{\textcolor[rgb]{0.84,0.33,0.22}{##1}}}
\expandafter\def\csname PY@tok@nl\endcsname{\let\PY@bf=\textbf\def\PY@tc##1{\textcolor[rgb]{0.00,0.13,0.44}{##1}}}
\expandafter\def\csname PY@tok@nn\endcsname{\let\PY@bf=\textbf\def\PY@tc##1{\textcolor[rgb]{0.05,0.52,0.71}{##1}}}
\expandafter\def\csname PY@tok@no\endcsname{\def\PY@tc##1{\textcolor[rgb]{0.38,0.68,0.84}{##1}}}
\expandafter\def\csname PY@tok@na\endcsname{\def\PY@tc##1{\textcolor[rgb]{0.25,0.44,0.63}{##1}}}
\expandafter\def\csname PY@tok@nb\endcsname{\def\PY@tc##1{\textcolor[rgb]{0.00,0.44,0.13}{##1}}}
\expandafter\def\csname PY@tok@nc\endcsname{\let\PY@bf=\textbf\def\PY@tc##1{\textcolor[rgb]{0.05,0.52,0.71}{##1}}}
\expandafter\def\csname PY@tok@nd\endcsname{\let\PY@bf=\textbf\def\PY@tc##1{\textcolor[rgb]{0.33,0.33,0.33}{##1}}}
\expandafter\def\csname PY@tok@ne\endcsname{\def\PY@tc##1{\textcolor[rgb]{0.00,0.44,0.13}{##1}}}
\expandafter\def\csname PY@tok@nf\endcsname{\def\PY@tc##1{\textcolor[rgb]{0.02,0.16,0.49}{##1}}}
\expandafter\def\csname PY@tok@si\endcsname{\let\PY@it=\textit\def\PY@tc##1{\textcolor[rgb]{0.44,0.63,0.82}{##1}}}
\expandafter\def\csname PY@tok@s2\endcsname{\def\PY@tc##1{\textcolor[rgb]{0.25,0.44,0.63}{##1}}}
\expandafter\def\csname PY@tok@vi\endcsname{\def\PY@tc##1{\textcolor[rgb]{0.73,0.38,0.84}{##1}}}
\expandafter\def\csname PY@tok@nt\endcsname{\let\PY@bf=\textbf\def\PY@tc##1{\textcolor[rgb]{0.02,0.16,0.45}{##1}}}
\expandafter\def\csname PY@tok@nv\endcsname{\def\PY@tc##1{\textcolor[rgb]{0.73,0.38,0.84}{##1}}}
\expandafter\def\csname PY@tok@s1\endcsname{\def\PY@tc##1{\textcolor[rgb]{0.25,0.44,0.63}{##1}}}
\expandafter\def\csname PY@tok@gp\endcsname{\let\PY@bf=\textbf\def\PY@tc##1{\textcolor[rgb]{0.78,0.36,0.04}{##1}}}
\expandafter\def\csname PY@tok@sh\endcsname{\def\PY@tc##1{\textcolor[rgb]{0.25,0.44,0.63}{##1}}}
\expandafter\def\csname PY@tok@ow\endcsname{\let\PY@bf=\textbf\def\PY@tc##1{\textcolor[rgb]{0.00,0.44,0.13}{##1}}}
\expandafter\def\csname PY@tok@sx\endcsname{\def\PY@tc##1{\textcolor[rgb]{0.78,0.36,0.04}{##1}}}
\expandafter\def\csname PY@tok@bp\endcsname{\def\PY@tc##1{\textcolor[rgb]{0.00,0.44,0.13}{##1}}}
\expandafter\def\csname PY@tok@c1\endcsname{\let\PY@it=\textit\def\PY@tc##1{\textcolor[rgb]{0.25,0.50,0.56}{##1}}}
\expandafter\def\csname PY@tok@kc\endcsname{\let\PY@bf=\textbf\def\PY@tc##1{\textcolor[rgb]{0.00,0.44,0.13}{##1}}}
\expandafter\def\csname PY@tok@c\endcsname{\let\PY@it=\textit\def\PY@tc##1{\textcolor[rgb]{0.25,0.50,0.56}{##1}}}
\expandafter\def\csname PY@tok@mf\endcsname{\def\PY@tc##1{\textcolor[rgb]{0.13,0.50,0.31}{##1}}}
\expandafter\def\csname PY@tok@err\endcsname{\def\PY@bc##1{\setlength{\fboxsep}{0pt}\fcolorbox[rgb]{1.00,0.00,0.00}{1,1,1}{\strut ##1}}}
\expandafter\def\csname PY@tok@kd\endcsname{\let\PY@bf=\textbf\def\PY@tc##1{\textcolor[rgb]{0.00,0.44,0.13}{##1}}}
\expandafter\def\csname PY@tok@ss\endcsname{\def\PY@tc##1{\textcolor[rgb]{0.32,0.47,0.09}{##1}}}
\expandafter\def\csname PY@tok@sr\endcsname{\def\PY@tc##1{\textcolor[rgb]{0.14,0.33,0.53}{##1}}}
\expandafter\def\csname PY@tok@mo\endcsname{\def\PY@tc##1{\textcolor[rgb]{0.13,0.50,0.31}{##1}}}
\expandafter\def\csname PY@tok@mi\endcsname{\def\PY@tc##1{\textcolor[rgb]{0.13,0.50,0.31}{##1}}}
\expandafter\def\csname PY@tok@kn\endcsname{\let\PY@bf=\textbf\def\PY@tc##1{\textcolor[rgb]{0.00,0.44,0.13}{##1}}}
\expandafter\def\csname PY@tok@o\endcsname{\def\PY@tc##1{\textcolor[rgb]{0.40,0.40,0.40}{##1}}}
\expandafter\def\csname PY@tok@kr\endcsname{\let\PY@bf=\textbf\def\PY@tc##1{\textcolor[rgb]{0.00,0.44,0.13}{##1}}}
\expandafter\def\csname PY@tok@s\endcsname{\def\PY@tc##1{\textcolor[rgb]{0.25,0.44,0.63}{##1}}}
\expandafter\def\csname PY@tok@kp\endcsname{\def\PY@tc##1{\textcolor[rgb]{0.00,0.44,0.13}{##1}}}
\expandafter\def\csname PY@tok@w\endcsname{\def\PY@tc##1{\textcolor[rgb]{0.73,0.73,0.73}{##1}}}
\expandafter\def\csname PY@tok@kt\endcsname{\def\PY@tc##1{\textcolor[rgb]{0.56,0.13,0.00}{##1}}}
\expandafter\def\csname PY@tok@sc\endcsname{\def\PY@tc##1{\textcolor[rgb]{0.25,0.44,0.63}{##1}}}
\expandafter\def\csname PY@tok@sb\endcsname{\def\PY@tc##1{\textcolor[rgb]{0.25,0.44,0.63}{##1}}}
\expandafter\def\csname PY@tok@k\endcsname{\let\PY@bf=\textbf\def\PY@tc##1{\textcolor[rgb]{0.00,0.44,0.13}{##1}}}
\expandafter\def\csname PY@tok@se\endcsname{\let\PY@bf=\textbf\def\PY@tc##1{\textcolor[rgb]{0.25,0.44,0.63}{##1}}}
\expandafter\def\csname PY@tok@sd\endcsname{\let\PY@it=\textit\def\PY@tc##1{\textcolor[rgb]{0.25,0.44,0.63}{##1}}}

\makeatother

\providecommand*{\DUrole}[2]{%
  \ifcsname DUrole#1\endcsname%
    \csname DUrole#1\endcsname{#2}%
  \else
    \ifcsname docutilsrole#1\endcsname%
      \csname docutilsrole#1\endcsname{#2}%
    \else%
      #2%
    \fi%
  \fi%
}

\ifthenelse{\isundefined{\hypersetup}}{
  \usepackage[colorlinks=true,linkcolor=blue,urlcolor=blue]{hyperref}
  \urlstyle{same} 
}{}

\begin{document}
\newcounter{footnotecounter}\title{SPySort: Neuronal Spike Sorting with Python}\author{Christophe Pouzat$^{\setcounter{footnotecounter}{1}\fnsymbol{footnotecounter}\setcounter{footnotecounter}{2}\fnsymbol{footnotecounter}}$%
          \setcounter{footnotecounter}{1}\thanks{\fnsymbol{footnotecounter} %
          Corresponding author: \protect\href{mailto:christophe.pouzat@parisdescartes.fr}{christophe.pouzat@parisdescartes.fr}}\setcounter{footnotecounter}{2}\thanks{\fnsymbol{footnotecounter} MAP5 lab., Paris-Descartes University and CNRS, Paris, France}, Georgios Is. Detorakis$^{\setcounter{footnotecounter}{3}\fnsymbol{footnotecounter}}$\setcounter{footnotecounter}{3}\thanks{\fnsymbol{footnotecounter} LSS, Supélec, Gif-sur-Yvette, France}\thanks{%

          \noindent%
          Copyright\,\copyright\,2014 Christophe Pouzat et al. This is an open-access article distributed under the terms of the Creative Commons Attribution License, which permits unrestricted use, distribution, and reproduction in any medium, provided the original author and source are credited. http://creativecommons.org/licenses/by/3.0/%
        }}\maketitle
          \renewcommand{\leftmark}{PROC. OF THE 7th EUR. CONF. ON PYTHON IN SCIENCE (EUROSCIPY 2014)}
          \renewcommand{\rightmark}{SPYSORT: NEURONAL SPIKE SORTING WITH PYTHON}

\setcounter{page}{27}
\newcommand*{\docutilsroleref}{\ref}
\newcommand*{\docutilsrolelabel}{\label}
\AtEndDocument{\cleardoublepage}
\begin{abstract}Extracellular recordings with multi-electrode arrays is one of the basic tools of contemporary neuroscience.
These recordings are mostly used to monitor the activities, understood as sequences of emitted action potentials,
of \emph{many} individual neurons. But the raw data produced by extracellular recordings are most commonly
a \emph{mixture} of activities from several neurons. In order to get the activities of the individual contributing
neurons, a pre-processing step called \emph{spike sorting} is required. We present here a pure Python implementation
of a well tested spike sorting procedure. The latter was designed in a modular way in order to favour a smooth
transition from an interactive sorting, for instance with IPython, to an automatic one. Surprisingly enough—or sadly enough,
depending on one's view point—, recoding our now 15 years old procedure into Python was the occasion of
major methodological improvements.\end{abstract}\begin{IEEEkeywords}clustering, sampling theorem, sampling jitter correction, dimension reduction, E-M algorithm, Gaussian Mixture Model, kmeans.\end{IEEEkeywords}

\section{Introduction%
  \label{introduction}%
}

The role of neuronal synchronisation in the information processing performed by (actual) neuronal networks is an actively debated question in neuroscience. Direct experimental measurement of synchronisation requires the recording of the activities of \textquotedbl{}as many neurons as possible\textquotedbl{} with a fine time resolution. In this context, \href{http://en.wikipedia.org/wiki/Multi-electrode_array}{multi-electrode arrays} (MEA) recordings constitute nowadays the technique of choice. The electrodes making an MEA are located in the extracellular space and can thereby record the action potentials or \emph{spikes} emitted by many neurons in their vicinity—an analogy is provided by a microphone for an electrode and many static people talking all at once in a language unknown to us for the neurons. Electrophysiologists can therefore monitor many neurons with a \textquotedbl{}limited\textquotedbl{} tissue damage—the more electrodes are pushed into a tissue, the more damage ensues: a very attractive feature of the methodology. However this attractive feature of multiple neurons recordings comes at a price: since many neurons are recorded from a single electrode, the raw data are a \emph{mixture} (of single neuron activities) and a \emph{comprehensive} use of the data requires the separation of this mixture into its individual components. This \textquotedbl{}separation\textquotedbl{} step is what is referred to as \emph{spike sorting} in neurophysiology.

Extracellular recordings have been used for a long time (60 years at least) and it is not surprising that many \emph{spike sorting} procedures have appeared in the literature (see \cite{Ein12} for a recent review). Extracellular recordings are also used daily in an applied context when neurologists perform \href{http://en.wikipedia.org/wiki/Electromyography}{electromyography}—extracellular recording from skeletal muscles where the \emph{recorded} action potentials are generated by the muscular cells, not by neurons—giving data and data analysis problems very similar to the ones we have presented so far. Very similar spike sorting methods have been developed in the former context (\emph{e.g.}, \cite{McG85}) but scientists working in the different contexts (\textquotedbl{}neurons\textquotedbl{} and \textquotedbl{}muscles\textquotedbl{}) do not seem to be aware that they have colleagues doing the same thing on slightly different data! We present here a rather \textquotedbl{}simple\textquotedbl{} approach (in the realm of the existing ones) which is the one we have used most often in the last 15 years. This approach was published in 2002 \cite{Pou02} and was successively \textquotedbl{}incarnated\textquotedbl{} using IGOR Pro (\href{http://www.wavemetrics.com/}{Wavemetrics}), \href{http://www.scilab.org/fr}{Scilab}, MATLAB (\href{http://www.mathworks.fr/products/matlab/}{MathWorks}), \href{http://www.r-project.org/}{R} and now \href{https://www.python.org/}{Python}. This work which was initially planed as a recoding of our present R code into Python was also the occasion to re-think some of the key steps of our procedure. This lead to a major improvement (also back ported to our R code) in the way a specific step, the \emph{sampling jitter estimation and correction}, is performed. This new development is given due space in section \DUrole{ref}{jitter-estimation} of the present manuscript.

This contribution is written with two generic readers in mind: scientific python users and neurophysiologists doing spike sorting. For the first \textquotedbl{}reader\textquotedbl{} we present another example of an actual scientific data analysis problem that is easily handled within the scientific Python ecosystem. The second reader is likely to perform spike sorting with a commercial software provided by one of the MEA amplifiers manufacturers. We do not want to claim that these software are necessarily bad, but it is our experience that when we deal with data sets from different preparations, it is extremely useful to be able to \emph{adapt} our method to the specific features of the data. For instance, when switching from the first olfactory relay of a locust (\emph{Schistocerca americana}) to the first olfactory relay of a cockroach (\emph{Periplaneta americana})—they have many different features \cite{Cha07}, we will start in a interactive mode, say with IPython, using the method previously developed for the locust, try out some alternative approaches at the key steps (spike detection, dimension reduction, clustering) before settling on a new procedure involving only few experiment specific parameters. The nature of the Python environment providing interactive development and leading to \textquotedbl{}black box\textquotedbl{} procedures is a clear advantage here. Doing this kind of method adaptation is hard, not to say impossible, with commercial solutions implementing a \textquotedbl{}one size fits all\textquotedbl{} approach. It moreover turns out that it is, nowadays, not that hard to implement the full sequence of steps required for spike sorting thanks to environments like Python that are both user friendly and computationally efficient. So we hope to motivate our second reader to give a try to open solutions giving access to \textquotedbl{}what's going on under the hood\textquotedbl{}. In addition we are advocates of the \emph{reproducible research} paradigm (\cite{Sto14} , \cite{Del12}) and an implementation of the latter requires accessibility of the code used for a published analysis.

Two versions of the source code are available. A \textquotedbl{}simple\textquotedbl{} one associated with a step-by-step tutorial
\url{http://xtof.perso.math.cnrs.fr/locust_sorting_python.html}, the source file (in emacs org mode) necessary to produce an extended version of the present document is available at the following address:
\url{http://xtof.perso.math.cnrs.fr/org/PouzatDetorakis2014.org}. An object-oriented implementation of the method is available on Github:
\url{https://github.com/gdetor/SPySort}. The
present methodological developments will be soon merged with \href{https://github.com/OpenElectrophy/OpenElectrophy}{OpenElectrophy}.

\section{Data properties%
  \label{data-properties}%
}

The data used for illustration here were recorded from the first olfactory relay, \emph{the antennal lobe}, of a locust (\emph{Schistocerca americana}). Recording setting and acquisition details are described in \cite{Pou02}. The data are available and can be dowloaded with:%
\begin{quote}\begin{verbatim}
from urllib.request import urlretrieve
data_names = ['Locust_' + str(i) + '.dat.gz'
              for i in range(1,5)]
data_src = ['http://xtof.disque.math.cnrs.fr/data/'
            + n for n in data_names]
[urlretrieve(data_src[i],data_names[i])
 for i in range(4)]
\end{verbatim}

\end{quote}
They were stored as floats coded on 64 bits and compressed with gnuzip. 20 seconds of data sampled at 15 kHz are contained in these files. Four files corresponding to the four electrodes or recording sites of a \emph{tetrode} (see Sec. \DUrole{ref}{why-tetrode}) are used. The first second of data from the four recording sites is shown next (Figure \DUrole{ref}{FirstSecondFig}).\begin{figure}[hbt]\noindent\makebox[\columnwidth][c]{\includegraphics[scale=0.75]{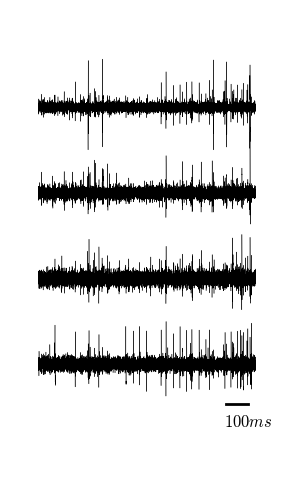}}
\caption{First second of data recorded from the four recording sites of a tetrode. \DUrole{label}{FirstSecondFig}}
\end{figure}

Here, the action potentials or spikes are the sharp (upward and downward) deviations standing out of the \textquotedbl{}noise\textquotedbl{}. When doing spike sorting we try to find \emph{how many different neurons} contribute to the data and, for each spike, what is the (most likely) neuron that generated it.

\subsection{Why tetrode?%
  \label{id8}%
  \label{why-tetrode}%
}

The main parameter controlling the amplitude of a recorded spike is the distance between the neuron and the electrode. It follows that if two similar neurons are equidistant to a given electrode, they will give rise to nearly identical spikes—for an elaboration on that and on how the signals recorded on different electrodes could be use to perform source localisation, see \cite{Che05}. These (nearly) identical recorded spikes are a big problem since the spike waveform (combination of shape and amplitude) is going to be our classification criterion. In some preparation, like the locust antennal lobe (but not the cockroach antennal lobe) using tetrodes, groups of four closely spaced electrodes, is going to help us as illustrated in figure \DUrole{ref}{WhyTetrodesFig}.\begin{figure}[htb]\noindent\makebox[\columnwidth][c]{\includegraphics[scale=0.75]{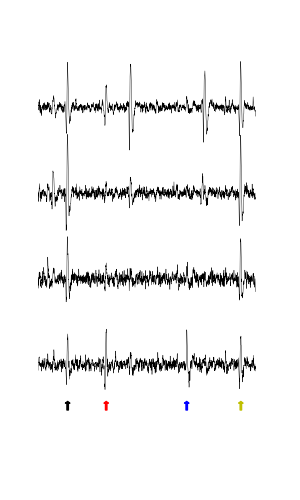}}
\caption{100 ms of data from the four recording sites of a tetrode. Four clear spikes on
the fourth recording site are marked by coloured arrows. \DUrole{label}{WhyTetrodesFig}}
\end{figure}

Imagine here that only the lowest electrode is available. Given the noise level, it would be hard to decide if the four spikes (arrows in figure \DUrole{ref}{WhyTetrodesFig}) are originating from the same neuron or not. If we now look at the same events from the additional viewpoints provided by the other electrodes (the three upper traces) it is clear that the four events cannot arise from the same neuron: the first and fourth events (seen on the lowest trace) are large on the four electrodes, while the second and third are large on the top and bottom traces but very tiny on the two middle traces.

\section{Main modelling assumptions%
  \label{main-modelling-assumptions}%
}

We will simplify the neurons discharge statistics by modelling them as independent Poisson processes—the successive inter spike intervals (ISI) of a given neuron are independently and identically distributed following an exponential distribution, they are also independent of the ISI of the other neurons. \emph{This is obviously a gross simplification}: we know that the ISI of a given neuron are not Poisson distributed and that the discharges of different neurons are correlated—that is precisely \emph{what we want to study with these experiments}—but the deviations of the actual data generation mechanism from our simple model (independent Poisson processes) has, in general, a negligible impact on the sorting results. If we want to work with more realistic models, we can (although not yet in Python), but the computational price is rather heavy (\cite{Pou04} and \cite{Del06}). We do go even further on the simplification path for these data since we are going to \textquotedbl{}forget\textquotedbl{} about the different discharge rates (at the classification stage, Sec. \DUrole{ref}{peeling}) and use only the amplitude information.

When a neuron fires a spike \emph{the same underlying waveform} with some additive auto-correlated Gaussian noise is recorded on each site (more precisely there is one waveform per electrode and per neuron). Four comments:%
\begin{itemize}

\item 

For some data sets (\emph{e.g.}, \cite{Del06}) the underlying waveform of a given neuron is changing during the discharge; we can model that if necessary (\cite{Pou04} and \cite{Del06}), but the computational cost is high and the neurons of the data set considered here do not exhibit this feature.
\item 

Following \cite{Che05} we could simplify the model assuming that we have a single \textquotedbl{}mother\textquotedbl{} waveform per neuron and that the underlying waveform seen on each electrode are just \emph{scaled} versions of the mother waveform. We haven't implemented this feature yet but it will come next.
\item 

Some authors \cite{Sho03} argue that the additive noise would be better described by a multivariate t-distribution; they are lead to this assumption because they do not resolve superposed events—when two or more neurons fire at nearly the same time the observed event is a \textquotedbl{}superposition\textquotedbl{}: the sum of the underlying waveforms of the different neurons plus noise. If superpositions are resolved, the Gaussian noise assumption is perfectly reasonable \cite{Pou02}.
\item 

The noise is necessarily auto-correlated since the data are low-pass filtered prior to digitisation.
\end{itemize}

\section{The sorting procedure%
  \label{the-sorting-procedure}%
  \label{sorting-procedure}%
}

A very detailed, \textquotedbl{}step-by-step\textquotedbl{}, account of the analysis presented here can be found on our dedicated web page (\url{http://xtof.perso.math.cnrs.fr/locust_sorting_python.html}). For most of the steps only a brief description is given in order to save space for the original part. We moreover focus on the first part of the analysis of what is typically a large data set. Experimentalists usually record for hours if not days \cite{Cha07} from the same preparation. In our experience such recordings are stable on a time scale of 10 minutes or more. It therefore makes perfect sense to split the analysis in two parts:\newcounter{listcnt0}
\begin{list}{\arabic{listcnt0}.}
{
\usecounter{listcnt0}
\setlength{\rightmargin}{\leftmargin}
}

\item 

Model estimation: in the \textquotedbl{}easy\textquotedbl{} settings as here, a model boils down to a catalogue of waveforms, one waveform per neuron and per recording site. More sophisticated models can be used but the case illustrated here—and \emph{that is not a rare case}—they are not necessary.
\item 

Once the model / waveform catalogue has been obtained the data are processed; that is events are detected and classification is performed by template matching—the catalogue's waveforms being the templates.\end{list}

The key point is that part 1 can be done on a short data stretch—in the example bellow we are going to use 10 seconds of data. This part is also the one that can require the largest amount of user input, in particular when a choice on the number of neurons to include in the model has to be made. The second part is straightforward to automate: a short Python script loading, say, 2 minutes of data and the catalogue will do the template matching as illustrated in Sec. \DUrole{ref}{peeling}. A \textquotedbl{}poor's man\textquotedbl{} illustration of this 2 parts approach is provided here since the model is estimated on the first half of the data set and the classification is performed on the whole set. When applying this approach, one should monitor the number of unclassified events over a given time period and \emph{update the model} if this number increases suddenly.

\subsection{Data normalisation%
  \label{id20}%
  \label{data-normalisation}%
}

If the data have not been high-passed filtered prior to digitization, they are so filtered (with a cutoff frequency between 200 and 500 Hz) using function \texttt{firwin} of module \href{http://docs.scipy.org/doc/scipy/reference/tutorial/signal.html\#fir-filter}{scipy.signal}.
The trace of each electrode is then \href{http://en.wikipedia.org/wiki/Median}{median} subtracted and divided by its \href{http://en.wikipedia.org/wiki/Median_absolute_deviation}{median absolute deviation} (MAD). The MAD provides a robust estimate of the standard deviation \emph{of the recording noise}. After this normalisation, detection thresholds are comparable on the different electrode.

\subsection{Spike detection%
  \label{id21}%
  \label{spike-detection}%
}

Spikes are detected as local extrema above a threshold. More precisely, the data are first filtered with a box filter (a moving average) in order to reduce the high frequency noise; the filtered data are normalised like the raw data before being \textquotedbl{}rectified\textquotedbl{}: amplitudes below a threshold are set to zero; the filtered and rectified data from each electrode are added together and local maxima are identified. This is a very simple method that works well for these data. This is clearly an important step that must typically be adapted to the data one works with. For instance when the signal to noise ratio is lower, we often construct a \textquotedbl{}typical waveform\textquotedbl{}—by detecting the largest events first, averaging and normalising them (peak at 1 and mean at 0)—that we convolve with the raw data. The detection is subsequently done on these filtered data. Working with an environment like Python we can do that with a few lines of code, try different ideas and different parameters, etc.

\subsection{Events set (sample) construction%
  \label{events-set-sample-construction}%
  \label{sample-construction}%
}

After a satisfying detection has been obtained, events are \textquotedbl{}cut\textquotedbl{} from the raw data. An optimal cut length is obtained by first using overly large cuts (say 80 sampling points on both sides of the detected peak). The point-wise MAD is computed and the locations at which the MAD reaches 1 (the noise level on the normalised traces) give the domain within which \textquotedbl{}useful sorting information\textquotedbl{} is to be found. New shorter cuts are then made (in the illustrated case, Fig.  \DUrole{ref}{First200Fig}, using 14 points before the peak and 30 points after) and an event is then described by a set of N amplitudes on 4 electrodes (in our case 180 amplitudes). The first 200 events are shown in Figure \DUrole{ref}{First200Fig}.\begin{figure}[hbt]\noindent\makebox[\columnwidth][c]{\includegraphics[scale=0.60]{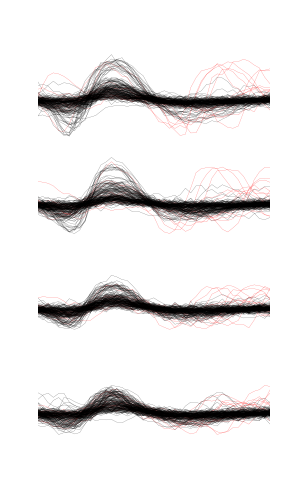}}
\caption{First 200 events: Black, non-superposed events; red, superpositions. The cuts are 3 ms (45 sampling points) long. Identical scales on each sub-plot. \DUrole{label}{First200Fig}}
\end{figure}

Superpositions (\emph{i.e.}, two or more spikes fired at nearly the same time by two or more neurons) are clearly visible as secondary peaks on each recording site (red in figure \DUrole{ref}{First200Fig}).

\subsection{Dimension reduction%
  \label{id22}%
  \label{dimension-reduction}%
}

The cuts shown in Fig. \DUrole{ref}{First200Fig} are 3 ms or 45 sampling points long. That means that our sample space has 45x4 = 180 dimensions. Our model assumptions imply that, in the absence of recording noise, each neuron would generate a single point in this space—strictly speaking, because of the sampling jitter (see Sec. \DUrole{ref}{jitter-estimation}), each neuron would generate a small cloud—and the recording noise will transform these \textquotedbl{}centers\textquotedbl{} into clouds, each cloud having the same variance-covariance matrix—this is of course expected only for the events that are not superpositions. At that stage sorting reduces to a \href{http://scikit-learn.org/stable/modules/clustering.html\#clustering}{clustering} problem and doing clustering in a 180 dimensional space is rarely a good idea. We therefore reduce the dimension of our events' space using principal component analysis (PCA) keeping only a few of the first principal components. But before that, the \textquotedbl{}most obvious\textquotedbl{} superpositions are removed from the sample. We do that because a few superpositions can dominate (and strongly corrupt) the result of a PCA analysis. The goal of this initial part of our procedure is moreover to build a catalogue of underlying waveform associated with each neuron. The actual sorting will be subsequently accounting for superpositions when they occur. The \textquotedbl{}most obvious superpositions\textquotedbl{} are removed by looking for side peaks on each individual event. Figure \DUrole{ref}{ScatMatFig} (made with \texttt{scatter\_matrix} of \href{http://pandas.pydata.org/}{pandas}) shows the events projected on the planes defined by every pair of the first four principal components.\begin{figure}[hbt]\noindent\makebox[\columnwidth][c]{\includegraphics[scale=0.75]{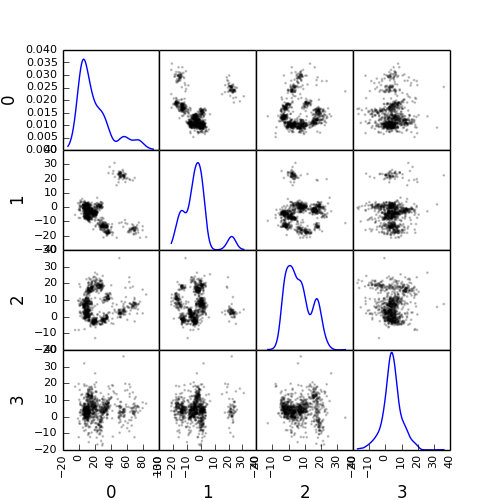}}
\caption{Scatter plot matrices of the events that are not superpositions on the plans defined by every pair of the first four principal components. \DUrole{label}{ScatMatFig}}
\end{figure}

We get an upper bound on the number of components to keep by building figures like Fig. \DUrole{ref}{ScatMatFig} with higher order components until the projected data look featureless (like a two dimensional Gaussian). We get an idea of the number of neurons by counting the number of clouds on the \textquotedbl{}good\textquotedbl{} projections (looking at the plot on row 1 and column 2 in Fig. \DUrole{ref}{ScatMatFig} we see 10 clouds).

\subsection{Dynamic visualisation%
  \label{id23}%
  \label{dynamic-visualisation}%
}

At that stage, dynamic visualisation can help a lot. We therefore typically export in \texttt{csv} format the data projected on the sub-space defined by principal components up to the upper bound found as just described. We then visualise the data with the free software \href{http://www.ggobi.org/}{GGobi}. The latter is extremely useful to: reduce further the dimension of the sub-space used; refine the initial guess on the number of clouds; evaluate the clouds shape (which conditions the clustering algorithm used).

\subsection{Clustering%
  \label{id24}%
  \label{clustering-kmeans}%
}

Although most of the spike sorting literature focuses on clustering methods, in our experience standard, well known and thoroughly tested methods work fine. After observing the data as in Fig. \DUrole{ref}{ScatMatFig} and with GGobi, we can decide what method should be used: a \textquotedbl{}simple\textquotedbl{} K-Means; a Gaussian mixture model (GMM) fitted with an E-M algorithm—both implemented in \href{http://scikit-learn.org/stable/}{scikit-learn}—; bagged-clustering \cite{Lei99} that we implemented in Python. For the data analysed here, we see 10 well separated clusters (clouds) that have uniform (spherical) shapes, suggesting that the K-Means are going to work well.

Figure \DUrole{ref}{FirstTwoClusters} shows the events attributed to the first 2 clusters. In order to facilitate model comparison (when models with different numbers of neurons are used or when a K-Means fit is compared with a GMM fit), clusters are ordered according to their centers' sizes. That is, for each cluster the point-wise median is computed and its size, the sum of its absolute values (an L1 norm), is obtained.\begin{figure}[hbt]\noindent\makebox[\columnwidth][c]{\includegraphics[scale=0.60]{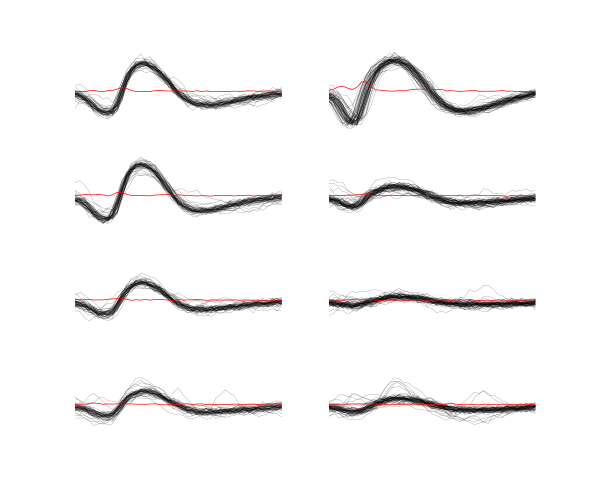}}
\caption{Left: the 52 events attributed to cluster 0. Right: the 65 events attributed to cluster 1. In red, the point-wise MAD (robust estimate of the standard deviation) \DUrole{label}{FirstTwoClusters}.}
\end{figure}

The point-wise MAD has been added to the events as a red trace in Fig. \DUrole{ref}{FirstTwoClusters}. If the reader remembers our modelling assumptions he or she will see a problem with the MAD of the second cluster (right column) on the top electrode: the MAD is clearly increasing on the rising phase of the event while our hypothesis imply that the MAD should be flat. But this MAD increase is obviously due to bad events' alignment. Seeing this kind of data, before rejecting our model hypothesis, we should try to better align the events to see if that could solve the problem. This is what we are going to do in the next section.

\subsection{Jitter estimation and cancellation%
  \label{jitter-estimation-and-cancellation}%
  \label{jitter-estimation}%
}

The \textquotedbl{}misaligned\textquotedbl{} events of Fig. \DUrole{ref}{FirstTwoClusters} (top right) have two origins. First, even in the absence of recording noise, we would have a jitter since the clock of our A/D card cannot be synchronised with the \textquotedbl{}clocks\textquotedbl{} of the neurons we are recording. This implies that when we are repetitively sampling spikes from a given neuron, the delay between the actual spike's peak and its closest sampling time \emph{will fluctuate} (in principle uniformly between -1/2 and +1/2 a sampling period). Since we are working with the sampled versions of the spikes and are aligning them on their apparent peaks, we are introducing a distortion or a \emph{sampling jitter} \cite{Pou02}. In addition, and that's the second origin of the misaligned events, we definitely have some recording noise present in the data and because of this noise we are going to make mistakes when we detect our local maxima at the very beginning of our procedure. In other words we would like to find local maxima of the \texttt{signal} but we can't do better (at that stage) than finding the local maxima of the \texttt{signal + noise}. Having a clear idea of the origin of the misalignment, we could decide that the MAD increase is not a real problem (we could in principle re-align the events and get rid of it) and live with it. Unfortunately, if we want to handle properly the superposed events, we have to estimate and compensate the sampling jitter as will soon become clear.

When we first published our method \cite{Pou02} we dealt with this jitter problem by using \href{http://en.wikipedia.org/wiki/Nyquist\%E2\%80\%93Shannon_sampling_theorem}{Nyquist theorem} that tells us that if our data were properly sampled—with a sampling frequency larger than twice the low-pass cutoff frequency of our acquisition filter—we can reconstruct \emph{exactly the data in-between our sampled points} by convolving the sampled data with a \texttt{sinc} function. We therefore went on, over sampling the data numerically, before shifting our individual events in order to align them on their cluster centre. This approach has several shortcomings: i) the support of the \texttt{sinc} is infinite but we are dealing with finite (in time) data and are therefore doing an approximate reconstruction; ii) computing the (approximate) interpolated values takes time. Luckily, recoding our procedure into Python led us to finally \textquotedbl{}see the light\textquotedbl{}—others \cite{Pil13} followed a similar path before us. We can indeed solve our problem much more efficiently, without using the \texttt{sinc} function.

Formally if we write $g(t)$, the observed waveform of an event within one of our cuts (the time \emph{t} runs from -1 ms to +2 ms in our examples), and $f(t)$, the underlying waveform—we are considering an event that is not a superposition and we write things for a single recording site to keep notations lighter, the generalisation to several recording sites is straightforward—we have:\begin{equation}
\label{jitter1}
g(t) = f(t+\delta) + Z(t) \, ,
\end{equation}where $\delta$ is the jitter we want to estimate and $Z(t)$ is a stationary and centred Gaussian process ($E(Z(t))=0$ and $\mathrm{Var}\left(Z(t)\right) = \sigma^2_Z$). Our approach seems to simplify considerably the estimation problem when compared to \cite{Pil13}. A second order Taylor expansion is used in our case, leading to:\begin{equation}
\label{jitter2}
g(t) \approx f(t) + \delta f'(t) + \delta^2/2 \, f''(t) + Z(t) \, .
\end{equation}If we assume that $\delta$ is the realisation of a random variable $\Delta$ with a null expectation, $\mathrm{E}(\Delta)=0$—that's a reasonable assumption given the origins of the jitter—and finite variance, $\sigma^2_{\Delta}$, then:\begin{equation}
\label{jitter3}
\mathrm{E}\left(g(t)\right) \approx f(t)  + \sigma^2_{\Delta}/2 \, f''(t) \, .
\end{equation}In other words, to the first order in $\delta$ (\emph{i.e.}, setting $\sigma^2_{\Delta}$ to 0), the expected value of the event equals the underlying waveform. Sticking to the first order we get for the variance:\begin{equation}
\label{jitter4}
\mathrm{Var}\left(g(t)\right) = \mathrm{E}\left[\left(g(t)-f(t)\right)^2\right] \approx  \sigma^2_{\Delta} \, f'(t)^2 + \sigma^2_Z \, .
\end{equation}Implying that the square root of the events variance minus the noise variance should be proportional to their absolute derivative; this explains why the MAD (a robust estimate of the standard deviation) peaks on the rising phase of the cluster centre (Fig. \DUrole{ref}{FirstTwoClusters}, top right) since that's where the time derivative is the largest.

Equation (\DUrole{ref}{jitter3}) tells us that our cluster centres estimated as point-wise median are likely to be \textquotedbl{}good\textquotedbl{} (in other words their error should be dominated by sampling variance, not by bias). Using the same argument, we can get first an estimate of the time derivative of the raw data by using the central difference (divided by two), then we can make cuts at the same locations and in exactly the same way as our original cuts and compute cluster specific point-wise medians giving us reasonable estimates of the time derivatives of the cluster centres (the $f'(t)$ above). We can iterate this procedure one step further to get estimates of the second derivatives of the cluster centres (the $f''(t)$ above).

We now have the required elements to go back to our jitter ($\delta$) estimation problem using Eq. (\DUrole{ref}{jitter2}). We don't have $g(t)$, $f(t)$, $f'(t)$ or $f''(t)$ directly but only sampled versions of those, that is: $\left[g_i=g(t_i)\right]_{i = t_1,\ldots,t_w}$, $\left[f_i=f(t_i)\right]_{i=t_1,\ldots,t_w}$ and $\left[f'_i=f'(t_i)\right]_{i=t_1,\ldots,t_w}$ where $w$ is the width of one of our cuts (45 sampling points). Starting with the first order in $\delta$, we can get an estimate $\tilde{\delta}$ of $\delta$ by minimising the residual sum of squares (RSS) criterion:\begin{equation}
\label{jitter5}
 \tilde{\delta} = \arg \min_{\delta} \sum_i \left(g_i - f_i - \delta \, f_i'\right)^2 \, .
\end{equation}Since the $(f_i)$ and $(f_i')$ are known, we are just solving a classical linear regression problem whose solution is:\begin{equation}
\label{jitter6}
 \tilde{\delta} = \frac{\sum_i (g_i - f_i) \,  f_i'}{\sum_i f_i'^2} \, .
\end{equation}We could take the noise auto-correlation (that we can estimate) into account, but it turns out to be not worth it (the precision gain is not really offsetting the computational cost).

We now solve the second order optimisation problem:\begin{equation}
\label{jitter7}
 \hat{\delta} = \arg \min_{\delta} \sum_i \left(g_i - f_i - \delta \, f_i' - \delta^2/2 \, f_i'' \right)^2 \, .
\end{equation}Since the latter does not admit (in general) a closed form solution, we perform a single \href{http://en.wikipedia.org/wiki/Newton-Raphson}{Newton-Raphson} step, starting from $\tilde{\delta}$ to get $\hat{\delta}$. Only a \emph{single} Newton-Raphson step is used because there is not much to be gained by refining the solution of an optimisation problem (Eq. \DUrole{ref}{jitter7}) that only provides an approximate solution to the problem we are really interested in—which would be written here: $\hat{\delta} = \arg \min_{\delta} \int \left(g(t)-f(t+\delta)\right)^2 dt$—the main error is likely to arise from the second order approximation of the latter—this point is clearly made in an other context, predictor-corrector method for ordinary differential equation, by Acton in \cite{Act70} on pp. 133-134.

Figure \DUrole{ref}{JitterCancellationIllustrated} illustrates jitter estimation and cancellation at work. The left column shows one of the events attributed to cluster 1 (black, $g(t)$ in our previous discussion) together with the cluster centre estimate (blue, $f(t)$ in our previous discussion) and the difference of the two (red,  $g(t)-f(t)$ in our previous discussion). The right column shows again the event (black) with the \emph{aligned} centre (blue, $f(t) + \hat{\delta} \, f'(t) + \hat{\delta}^2/2 \, f'^2(t)$ in the previous discussion) and the difference of the two (red).\begin{figure}[hbt]\noindent\makebox[\columnwidth][c]{\includegraphics[scale=0.60]{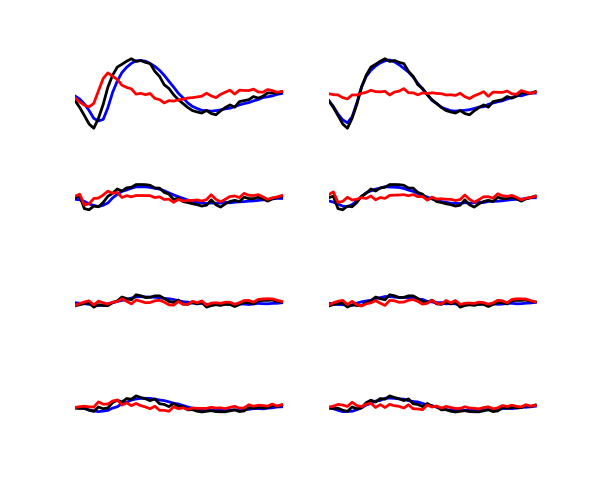}}
\caption{Left: event 50 of cluster 1 (black), centre of cluster 1 (blue), difference of the 2 (red). Right: event 50 of cluster 1 (black), \emph{aligned} centre of cluster 1 (blue), difference of the 2 (red) \DUrole{label}{JitterCancellationIllustrated}.}
\end{figure}

\subsection{Spikes \textquotedbl{}peeling\textquotedbl{}%
  \label{spikes-peeling}%
  \label{peeling}%
}
\begin{figure*}[]\noindent\makebox[\textwidth][c]{\includegraphics[scale=0.25]{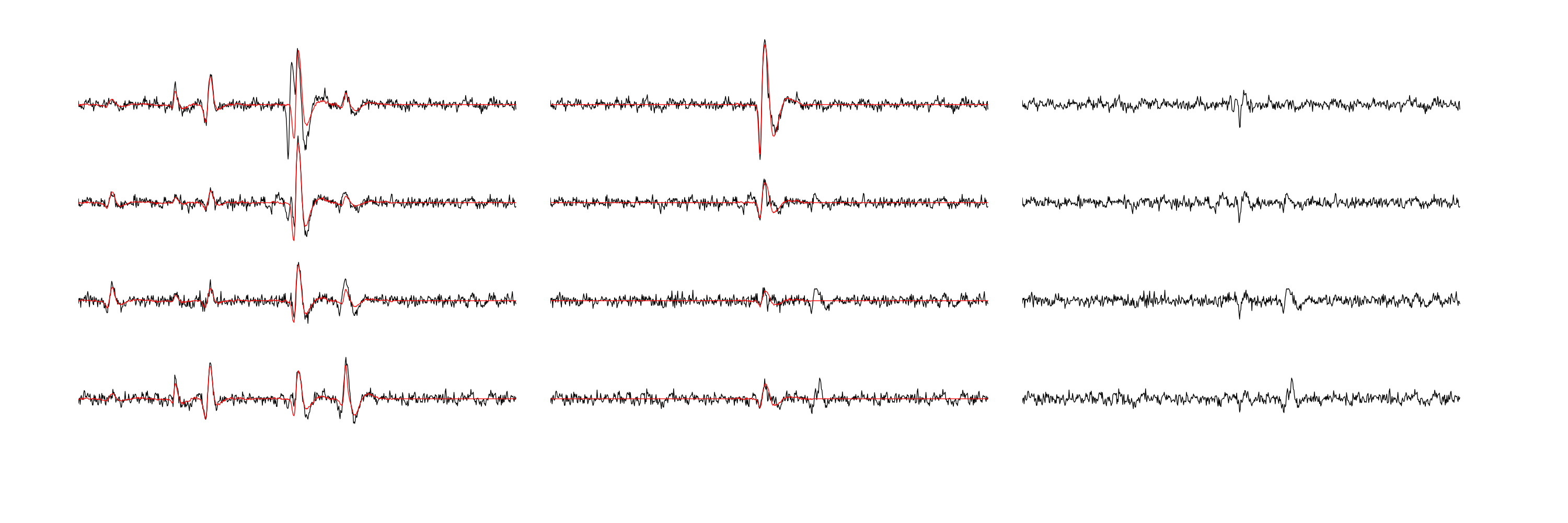}}
\caption{Illustration the \textquotedbl{}peeling\textquotedbl{} procedure. Left: raw data (black) and first prediction (red); middle: previous raw data minus previous prediction (black) and new prediction (red); right: what's left (no more waveforms corresponding to the catalogue's content). The small spike left on the right (clearly visible in the middle on the four sites) does not belong to any neuron of the catalogue because the events used to built the latter where detected as local maxima (and we would need to detect local minima to catch events like the one we see here) \DUrole{label}{PeelingIllustrated}.}
\end{figure*}

We have almost reached the end of our journey. The clustering step gave us a catalogue of waveforms: the cluster centre, its first and second derivative for each of the $K$ neurons / clusters on each site. We now go back to the raw data and for each detected event we do:\setcounter{listcnt0}{0}
\begin{list}{\arabic{listcnt0}.}
{
\usecounter{listcnt0}
\setlength{\rightmargin}{\leftmargin}
}

\item 

Compute the squared Euclidean norm of event (over the 4 cuts corresponding to the 4 electrodes) to get $R^2$.
\item 

For each of the $K$ neurons, align the centre's waveform on the event (as described in the previous section) and subtract it from the event. Compute the squared norm of this residual to get $R_j^2$ where $j=1,\ldots,K$.
\item 

Find $\hat{j} =\arg \min_j R_j^2$ and if $R_{\hat{j}}^2 < R^2$ then:%
\begin{itemize}

\item 

Keep the jitter corrected time for $\hat{j}$ in the list of spikes and keep $\hat{j}$ as the neuron of origin.
\item 

Subtract the $\hat{j}$-th aligned centre from the raw data
\end{itemize}

otherwise \emph{tag the event as unclassified} and don't perform any subtraction.\end{list}

Once every detected event has been examined, we are left with a \textquotedbl{}new\textquotedbl{} version of the raw data from which the aligned \textquotedbl{}best\textquotedbl{} centre waveforms have been subtracted (only when doing so was reducing the sum of squares of the amplitudes over the cuts). For the event illustrated in Fig. \DUrole{ref}{JitterCancellationIllustrated} we go from the black trace on the left column to the red trace on the right column. It is clear that for this \textquotedbl{}peeling procedure\textquotedbl{} to work we have to cancel the jitter otherwise we would be going from the black trace on the left column to the red trace \emph{on the same column} (where what remains as a peak amplitude similar to what we started with!).

We then iterate the procedure, taking the \textquotedbl{}new\textquotedbl{} raw data as if they were original data, detecting events as on the raw data, etc. We do that until we do not find anymore events for which the proposed subtraction is accepted; that is until we are only left with unclassified events. The first two iterations of this procedure are illustrated on figure \DUrole{ref}{PeelingIllustrated}. See how the superposed event in the middle of the trace (left column) is nicely resolved into its two components.

\section{Conclusions%
  \label{conclusions}%
}

Recoding our procedure from R to Python turned out to be easy (and an excellent way to learn Python for the first author). The efficient memory management provided by \texttt{numpy} for large arrays turns out to be very attractive. The \textquotedbl{}idiosyncrasies\textquotedbl{} of \texttt{matplotlib} (\emph{e.g.}, linewidth abbreviation is \texttt{lw} in Matplotlib and
\texttt{lwd} in R, color abbreviation is \texttt{c} in Matplotlib and in R is \texttt{col}, etc) turn out to be the longest to digest—for an R user—, but once they are mastered, IPython provides an excellent environment for interactive sorting. We are clearly going to carry out the subsequent developments of our methods—starting by porting our C code dealing with more sophisticated data generation models \cite{Pou04} and \cite{Del06} within the Python ecosystem.

More fundamentally, the new jitter estimation and cancellation procedure we introduced is deceptively simple—similar to the method of \cite{Pil13} but much simpler; to be fair, these authors also considered a possible amplitude and duration variability of the spikes generated by a given neuron. Our method is in fact, we think, an important step forward since it allows electrophysiologists to process superposed events systematically and \emph{efficiently}. And, in our view, without superposed events processing there is no trustworthy spike sorting.








\section{Acknowledgments%
  \label{acknowledgments}%
}

This work has been supported by the ANR JCJC project “SynchNeuro”.

\end{document}